\documentclass[useAMS, usenatbib]{mn2e}
\usepackage{amsmath}
\usepackage{xspace}
\usepackage{psfig}
\usepackage{defs}

\def\ie{{\it i.e.}}
\def\eg{{\it e.g.}}
\def\ltsima{$\; \buildrel < \over \sim \;$}
\def\simlt{\lower.5ex\hbox{\ltsima}}
\def\gtsima{$\; \buildrel > \over \sim \;$}
\def\simgt{\lower.5ex\hbox{\gtsima}}
\def\fesc{{$\langle f_{esc}\rangle$}\xspace}

\def\h2{H$_2$\xspace}
\def\ion#1#2{\text{#1\,\sc #2}}
\def\HI{{\ion{H}{i} }}
\def\HII{{\ion{H}{ii} }}

\title
{Did Globular Clusters Reionize the Universe?}
\author[M. Ricotti]
{Massimo Ricotti\\ 
Center for Astrophysics and Space Astronomy\\ University of
Colorado, Boulder, CO 80309\\ ricotti@casa.colorado.edu}
\date{Accepted ---. Received ---; in original form 15 March 2002}
\pagerange{\pageref{firstpage}--\pageref{lastpage}}
\pubyear{2002}
\begin{document}
\maketitle
\label{firstpage}

\begin{abstract}
  A problem still unsolved in cosmology is the identification of the
  sources of radiation able to reionize \HI in the intergalactic
  medium (IGM) by $z \sim 6$.  Theoretical works and observations seem
  to indicate that the fraction, \fesc, of \HI ionizing radiation
  emitted from galaxies that escapes into the IGM is small in the
  local universe (\fesc$\simlt 10$\%). At high redshift galaxies are
  more compact and probably gas rich implying smaller values of \fesc
  from their disks or spheroids. But if the sites of star formation
  are displaced from the disk or spheroid and the star formation
  efficiency of the proto-clusters is high, then \fesc should be about
  one.  This star formation scenario is consistent with several models
  for globular clusters formation. Using simple arguments based on the
  observed number of globular cluster systems in the local universe
  and assuming that the oldest globular clusters formed before
  reionization and had \fesc$\sim 1$, I show that they produced enough
  ionizing photons to reionize the IGM at $z \approx 6$.
\end{abstract}
\begin{keywords}
Globular clusters: general -- Galaxies: formation -- Cosmology: theory 
\end{keywords}

\section{Introduction and Rationale}

Observation of Ly$\alpha$ absorption systems toward newly found
high-redshift quasars \citep{Becker:01, Djorgovski:01}
indicate that the redshift of reionization of the intergalactic medium
(IGM) should be close to $z=6$ \citep{Gnedin:02, Songaila:02}.
Perhaps the recent identification of a lensed galaxy at $z=6.56$
points to a somewhat earlier redshift of reionization \citep{Hu:02}.
Although quasars play a dominant role in photoionizing the IGM at $z
\approx 3$ \citep{Meiksin:93}, their dwindling numbers at $z > 4$
suggest the need for another ionization source.  Unless a hidden
population of quasars is found, radiation emitted by high-redshift
massive stars seems necessary to reionize the universe.
A key ingredient in determining the effectiveness by which galaxies
photoionize the surrounding IGM is the parameter \fesc, defined here
as the mean fraction of Lyc photons escaping from galaxy halos into
the IGM. To be an important source of ionizing photons and rival with
quasars, a substantial fraction ($\sim 10$\%) of them must escape the
gas layers of the galaxies \citep{Madau:96}.

Cosmological simulations and semi-analytical models of IGM
reionization by stellar sources find that the ionizing background
rises steeply at the redshift of reionization. Unfortunately a direct
comparison between models is difficult because of different recipes
used for star formation, clumping of the IGM or the definition of
\fesc. But a result common to all the models is that, in order to
reionize the IGM by $z =6 - 7$, the escape fraction must be relatively
large: \fesc$\simgt 10$\% assuming a Salpeter initial mass function
(IMF) and the standard $\Lambda$CDM cosmological model.
\cite{Benson:02} finds that \fesc should be about 15\% for
reionization at $z = 6$, but a smaller value \fesc$ \simlt 10$\% is
consistent with the observed ionizing background at $z \sim 3$.
\cite{Gnedin:02} finds that assuming a primordial power spectrum index
$n=0.93$, the preferred value from CMB and LSS data, reionization at
$z \simgt 6$ requires a large \fesc; but this assumption produces an
ionizing background at $z \simlt 4$, that is too large.
The common assumption of a universal star formation efficiency
(SFE) (\eg, the coefficient in front of the Schmidt-Law in some models
or in others the fraction $f_*$ of baryons converted into stars) is
consistent with the observed values of the star formation rate (SFR)
at $0<z<5$ and total star fraction $\Omega_*$ at $z =0$. However,
the assumption of a constant \fesc does not seem to be consistent with
observations.  An escape fraction \fesc$\sim 1$ is required for
reionization at $z \sim 6$ but the ionizing background at $z \sim 3$
is consistent with \fesc$\simlt 10$\% \citep{Bianchi:01}.  Small
values of \fesc at $z \simlt 3$ are also supported by direct
observations of the Lyc emission from Lyman-break and starburst
galaxies.  \cite{Giallongo:02} find an upper limit \fesc$< 16$\% at $z
\sim 3$ \citep[but see][]{Steidel:01} and observations of low-redshift
starbursts are consistent with \fesc upper limits ranging from a few
percent up to 10\% \citep{Hurwitz:97,Deharveng:01}.

Calculations of \fesc from first principles are difficult. The main
complications arise in simulating a realistic interstellar medium
(ISM) that includes small-scale physics and feedback processes.
Moreover the mean \fesc results from the contribution of a variety of
galaxies whose ISM properties are largely unknown at high redshift.
Theoretical models \citep{Dove:00, Ciardi:02} for the radiative
transfer of ionizing radiation through the disk layer of spiral
galaxies similar to the Milky Way find \fesc$\sim 6-10$\%. At high
redshift the mean value of \fesc is expected to decrease almost
exponentially with increasing redshift \citep{RicottiS:00, Wood:00};
at $z > 6$, \fesc$\simlt 0.1-1$\% even assuming star formation rates
typical of starburst galaxies (\eg, SFR$\sim 10$ times that of the
Milky Way).  Using Monte-Carlo simulations, \cite{RicottiS:00} have
studied how \fesc depends on galactic parameters.  Assuming gas
density profile in hydrostatic equilibrium in the dark matter (DM)
potential, star density proportional to the gas density and a power
law for the luminosity function of the OB associations,
they found that \fesc$\propto(\epsilon f_g
M_{DM})^{-1/3}\exp[-(z_{vir}+1) \epsilon^{-1/3}]$. Here $\epsilon$ is
proportional to the SFE, $f_g$ is the fraction of collapsed gas,
$z_{vir}$ is the virialization redshift and $M_{DM}$ is the DM halo
mass. The majority of photons that escape the halo come from the most
luminous OB associations located in the outermost parts of the galaxy.
Indeed \cite{RicottiS:00} have shown that changing the luminosity
function of the OB association and the density distribution of the
stars has major effects on \fesc (see their Figs. 8 and 9). In the
aforementioned models, \fesc should be regarded as an upper limit
since dust extinction and absorption of ionizing radiation from the
molecular cloud in which OB associations are born are neglected.

The theoretical suggestion of a decreasing \fesc with increasing
redshift is in contrast with models for reionization that require
\fesc$\sim 1$ at $z = 6$. A different star formation mode, with very
luminous OB associations forming in the outer parts of galaxy halos,
could explain the large \fesc required for reionization. Globular
clusters (GCs) are possible observable relics of such a star formation
mode.  Their redshift of formation is compatible with redshift of
reionization \citep{GnedinR:02}.  Because of their large star density
they survived tidal destruction and represent the most luminous tail
of the luminosity distribution of primordial OB associations.  In
\S~\ref{ssec:mods} I explain that several models for the formation of
proto-GCs imply an \fesc$\sim 1$.  I will also show that the total
amount of stars in GCs observed today is sufficient to reionize the
universe at $z \sim 6$ if their \fesc$\sim 1$. This conclusion is
reinforced if the GCs we observe today are only a fraction,
$1/f_{di}$, of primordial GCs as a consequence of mass segregation and
tidal stripping.

The paper is organized as follows. In \S~\ref{sec:rev} I briefly
review recent progress in our understanding of GC properties and
formation theories; in \S~\ref{sec:meth} I discuss the model
assumptions in light of GC observations and present the
results.  In \S~\ref{sec:disc} I present my conclusions.

\section{Short review on GC systems}\label{sec:rev}

In this section I review some observational and theoretical results
on GCs useful to the aim of this paper. I also try to justify my
assumption \fesc$\sim 1$ for GCs on the base of theoretical models of
proto-GC formation. 
  
\subsection{Observations}\label{ssec:obs}

Most galaxies have a bimodal GCs distribution indicating that luminous
galaxies experience at least two major episodes of GCs formation. The
bulk of the globulars in the main body of the Galactic halo appear to
have formed during a short-lived burst ($\sim 0.5-2$ Gyr) that took
place about 13 Gyrs ago. This was followed by a second burst
associated with the formation of the galactic bulges.  Clusters may
have been formed in dwarf spheroidal galaxies and acreted by the
Galactic halo \citep{vandenBergh:99}. Massive cluster formation
occurred in galaxies as small as the Fornax dwarf spheroidal but not
in massive ones such as the small magellanic cloud \citep{Zepf:99}.

\subsubsection{Absolute and relative ages}\label{sssec:age}

The method for determinating the absolute age of GCs is based on
fitting the observed color-magnitude diagram with theoretical
evolutionary tracks. The systematics in the evolutionary model and the
determination of the cluster distance are the major sources of errors.
Recent determinations of the absolute age of old GCs find $t_{gc} =
12.5 \pm 1.2$ Gyr \citep{Chaboyer:98}, consistent with
radioactive dating of a very metal-poor star in the halo of our galaxy
\citep{Cayrel:01}.  Relative ages of Galactic GCs can be determined
with greater accuracy, since many systematic errors can be eliminated.
In our Galaxy $\Delta t_{gc}=0.5$ Gyr, but differences in age between
GC systems in different galaxies could be $\Delta t_{gc} \sim 2$ Gyr
\citep{Stetson:96}.

\subsubsection{Specific frequency}\label{sssec:sf}

The specific frequency is defined as the number, $N$, of GCs per
$M_V=-15$ of parent galaxy light, $S_N = N \times 10^{0.4(M_V+15)}$
(Harris \& van den Bergh 1981). The most striking characteristic is
that $S_N({\rm Ellipticals})>S_N({\rm Spirals})$.  $S_N = 0.5$ in
Sc/Ir galaxies (Harris, 1991), $S_N = 1$ in spirals of types Sa/Sb ,
$S_N =2.5$ in field ellipticals \citep{KunduI:01, KunduII:01}.
Converting to luminosity ($L_V/L_\odot=10^{-0.4(M_V-4.83)}$) we have
$N = (L/L_\odot)S_N/8.55 \times 10^7$. I can therefore calculate the
efficiency of GC formation defined as,
\begin{equation}
\epsilon_{gc}={M_{gc} \over M}={f_{di}Nm_{gc} \over M}={S_N f_{di} \over (M/L)_V}
\times 0.00585,
\label{eq:ef}
\end{equation}
where $M_{gc}$ is the total mass of the GC system, $m_{gc}=5 \times
10^5$ M$_\odot$ is the mean mass of GCs today, $M$ is the stellar
mass and $(M/L)_V$ is the mass to light ratio of the galaxy. In the
next paragraph we show that, because of dynamical evolution, $m_{gc}$
and $N$ are expected to be larger at the time of GC formation than
today.  Therefore, the parameter $f_{di} \ge 1$ is introduced to
account for dynamical disruption of GCs during their lifetime.

\subsubsection{IMF, Metallicity and Dynamical Evolution}\label{sssec:imf}

The IMF of GCs is not known.  The present mass function is known only
between 0.2 and 0.8 M$_\odot$ since high-mass stars are lost because of
two-body relaxation and stellar evolution processes.
Theoretical models show that the shape is consistent with a Salpeter-like
IMF.  The mean metallicity of old GCs is $Z \sim 0.03~Z_\odot$. One of
the most remarkable properties of GCs is the uniformity of their
internal metallicity $\Delta[Fe/H] \simlt 0.1$. This implies that the
bulk of the stars that constitute a GC formed in a single monolithic
burst of star formation. A typical GC emits $S \approx 3 \times
10^{53}$ s$^{-1}$ ionizing photons in a burst lasting 4 Myrs: about
$300$ times the ionizing luminosity of largest OB associations in our
Galaxy.

During their lifetime GCs lose a large part of their initial mass or
are completely destroyed by internal and external processes.
Prominent internal processes are mass loss by stellar evolution (about
half of the initial mass is lost) and two-body relaxation, the effects
of which are mass segregation (change in the mass function) and core
collapse (expansion and evaporation). External processes can be
divided roughly into two classes: gravitational shocks from GC motion
through the disk or bulge of the galaxy, and tidal forces, which cause
mass loss due to tidal truncation. Numerical simulations show that
about 50\%--90\% of the mass of GCs is lost due to external processes,
depending on the host galaxy environment, initial concentration and
IMF of the proto-GCs \citep{Chernoff:90, GnedinO:97}. Many of the
low-metallicity halo field stars in the Milky-Way could be debris of
disrupted GCs. The mass in stars in the halo is about 100 times the
mass in GCs. Therefore the parameter $f_{di}$, defined in
\S~\ref{sssec:sf}, could be as large as $f_{di}=100$. Overall $f_{di}$
is not well constrained since it depends on unknown properties of the
proto-GCs. According to results of N-body simulations $f_{di}$ should
be in the $f_{di} \sim 2-10$ range.

\subsection{Why is \fesc$\sim 1$ plausible for GCs?}\label{ssec:mods}

I discuss separately two issues: (i) the
\fesc from the gas cloud in which the GC forms, and (ii) the \fesc
through any surrounding gas in the galaxy.
\begin{itemize}
\item[\bf{(i)}] The evidence for \fesc$\sim 1$ comes from the observed
properties of present-day GCs. The fact that they are compact
self-gravitating systems with low and uniform metallicity points to a
high efficiency of conversion of gas into stars. A longer timescale of
star formation would have enriched the gas of metals and the
mechanical feedback from SN explosions would have stopped further star
formation. If $f_*\approx 10$\% of the gas is converted into stars in
a single burst (with duration $< 4$ Myr) at the center of a spheroidal
galaxy, following the simple calculations shown in \cite{RicottiS:00}
[see their eq.~(18)] at $z=6$ we have,
\begin{equation}
  f_{esc}=1-0.06{(1-f_*)^2\over f_*}\left({1+z \over 7}\right)^3\sim 50\%.
\label{eq:fc}
\end{equation}
\item[{\bf(ii)}] The justification for \fesc$\sim 1$ is model
  dependent but in general there are two main arguments: a) the high
  efficiency of star formation $f_*$, and b) the sites of proto-GC
  formation in the outermost parts of the galaxy halo.
  
  In the {\it ``cosmological objects model''} ($30<z_f<7$) of
  \cite{Peebles:68} GCs form with efficiency $f_* \approx 100\%$,
  implying \fesc$=1$ (note that such a high $f_*$ is not found in
  numerical simulations of first object formation
  \citep{RicottiGSa:02, RicottiGSb:02}). In {\it ``hierarchical
  formation models''} ($10<z_f<3$) \citep{Larson:93,Harris:94,
  McLaughlin:96, GnedinR:02} GCs form in the disk or spheroid of
  galaxies with gas mass $M_g \sim 10^7-10^9$ M$_\odot$. Compact GCs
  survive the accretion by larger galaxies while the rest of the
  galaxy is tidally stripped. Assuming that $1-10$ GCs form in a
  galaxy with $M_g \sim 10^7-10^8$ M$_\odot$ implies $f_* \sim 10\%$
  and therefore \fesc$\ge 50\%$. \fesc is larger than in
  eq.~(\ref{eq:fc}) if proto-GCs are located off-center (\eg, if they
  form from cloud-cloud collisions during the galaxy assembly) or if
  part of the gas in the halo is collisionally ionized as a
  consequence of the virialization process.\\ In models such as the
  {\it ``supershell fragmentation''} ($z_f<10$) of \cite{Taniguchi:99} or
  the {\it ``thermal instability''} ($z_f<7$) of \cite{Fall:85},
  \fesc$\approx 1$ since proto-GCs form in the outermost part of an
  already collisionally ionized halo.
\end{itemize}
In summary, since \fesc depends strongly on the luminosity of the OB
  associations and on their location, proto-GCs, being several hundred
  times more luminous than Galactic OB associations, should have a
  comparably larger \fesc. 

\section{Method and Results}\label{sec:meth}

In this section I estimate the number of ionizing photons emitted per
baryon per Hubble time, ${\cal N}_{ph}$, by GC formation. In
\S~\ref{ssec:simp} I derive ${\cal N}_{ph}$ assuming that all GCs
observed at $z=0$ formed in a time period $\Delta t_{gc}$ with
constant formation rate.  In \S~\ref{ssec:ps} I use the
Press-Schechter formalism to model more realistically the formation
rate of old GCs.
\begin{table}
\centering
\caption{Star census at z=0.}\label{tab:one}
\begin{tabular*}{8.3 cm}[]{l|c|c|c|c}
Type & $\omega_*$ (\%) & $S_N$ & (M/L)$_V$ & $\epsilon_{gc}$ (\%)\\
\hline
Sph & $6.5_{-3}^{+4}$ & $2.4 \pm 0.4$ & $5.4 \pm 0.3$ & $0.26\pm 0.06$\\
Disk & $2_{-0.5}^{+1.5}$ & $1 \pm 0.1$ & $1.82 \pm 0.4$ & $0.32\pm 0.1$ \\
Irr & $0.15_{-0.05}^{+0.15}$ & 0.5 & $1.33 \pm 0.25$ & $0.22\pm 0.04$\\
\hline
Total & $9_{-3.5}^{+5.5}$ & - & - &  $0.3\pm 0.07$\\
\end{tabular*}
\end{table}

\subsection{The simplest estimate}\label{ssec:simp}

I start by estimating the fraction, $\omega_{gc}$, of cosmic baryons
converted into GC stars. By definition
$\omega_{gc}=\omega_*\epsilon_{gc}$, where $\omega_*$ is the fraction
of baryons in stars at $z=0$ and $\epsilon_{gc}$ is the efficiency of
GC formation defined in \S~\ref{sssec:sf}. In all the calculations I
assume $\Omega_b=0.04$.  In Table~\ref{tab:one}, I summarize the star
census at $z=0$ according to \cite{Fukugita:98} and I derive
$\epsilon_{gc}$ using eq.~(\ref{eq:ef}), assuming $f_{di}=1$. Using
similar arguments \cite{McLaughlin:99} finds a universal efficiency of
globular cluster formation $\epsilon_{gc}=(0.26 \pm 0.05)$\%, in
agreement with the simpler estimate presented here.  It follows that
$\omega_{gc}=f_{di}(2.7^{+2.3}_{-1.7} \times 10^{-4})$ at $z=0$.

The total number of ionizing photons per unit time emitted by GCs is
$\eta\omega^f_{gc}/\Delta t_{gc}$, where $\eta$ is the number of
ionizing photons emitted per baryon converted into stars, and
$\omega^f_{gc} \approx 2.1 \omega_{gc}$ takes into account the mass
loss due to stellar winds and SN explosions adopting an
instantaneous-burst star formation law. GCs did not recycle this lost
mass since they formed in a single burst of star formation.  $\eta$
depends on the IMF and on the metallicity of the star. I calculate
$\eta$ using a Salpeter IMF and star metallicity $Z =0.03~Z_\odot$
(see \S~\ref{sssec:imf}) with Starburst99 code \citep{Leitherer:99},
and find $\eta=8967$.
The number of ionizing photons per baryon emitted in a Hubble
time at $z=6$ is,
\begin{equation}
{\cal N}^{gc}_{ph}=\eta \omega^f_{gc} {t_H(z=6) \over \Delta t_{gc}}=(5.1^{+4.3}_{-3.2}){f_{di} \over \Delta t_{gc}~(Gyr)},
\label{eq:ns}
\end{equation}
where I have assumed \fesc$=1$ and Hubble time at $z=6$ $t_H=1\pm 0.1$
Gyr. I expect $1 \le f_{di} \simlt 100$ and $0.5 \simlt \Delta t_{gc}
\simlt 2$ Gyr. A conservative estimate of $f_{di} \simgt 2$ and
$\Delta t_{gc} \simlt 2$ Gyr (\ie, $10 < z_f < 3$) implies
$f_{di}/\Delta t_{gc} \simgt 1$. The IGM is reionized when ${\cal
N}_{ph}=C$, where $C=\langle n_{HII}^2 \rangle/\langle n_{HII}
\rangle^2$ is the ionized IGM clumping factor. According to the
adopted definition of \fesc, $C=1$ for a homogeneous IGM, or $1 \simlt
C \simlt 10$ taking into account IGM density fluctuations producing
the Ly$\alpha$ forest \citep{Miralda:00, Gnedin:00}. The estimate from
eq.~(\ref{eq:ns}) is rather rough because I have implicitly assumed
that the SFR is constant during the period of GC formation $\Delta
t_{gc}$. A more realistic SFR as a function of redshift requires
assuming a specific model for the formation of GCs.  I try to address
this question in the next section.

\subsection{A bit of modeling}\label{ssec:ps}

\begin{figure}
\centerline{\psfig{figure=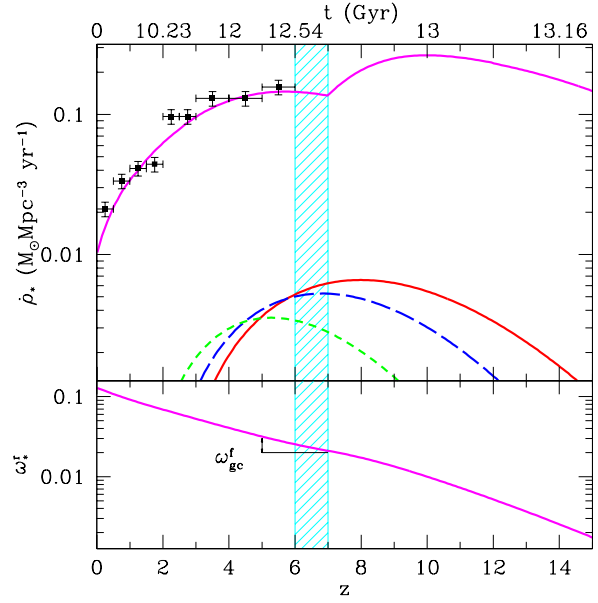,height=8cm}}
\caption{\label{fig:sfr} The solid line in the top panel shows the
  comoving SFR as a function of time in our model. The solid, dashed
  and short-dashed lines show the SFR of GCs for cases (i), (ii) and
  (iii), respectively (assuming $f_{di}=2$). The bottom panel shows
  $\omega^f_*$ as a function of time. The segment with arrows is a
  visual aid to compare $\omega^f_{gc}$ (assuming $f_{di}=20$) to
  $\omega^f_*$ around $z \sim 6$.}
\end{figure}

\begin{figure}
\centerline{\psfig{figure=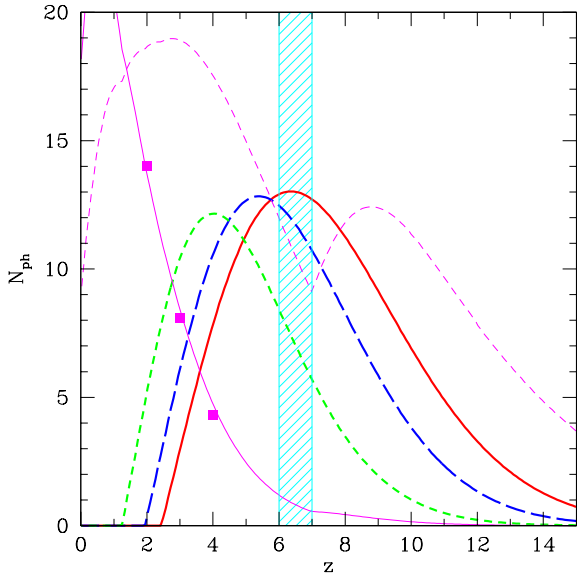,height=8cm}}
\caption{\label{fig:nph} Emissivity (photons per baryon per Hubble
  time) as a function of redshift. The thick solid, dashed and
  short-dashed lines show the contribution of GCs for cases (i), (ii)
  and (iii), respectively (assuming $f_{di}=2$).  The thin lines show
  the contribution of galaxies assuming a realistic \fesc$=0.1 \times
  \exp[-z/2]$ (solid) and a constant \fesc$=5$\% (dashed).}
\end{figure}

I assume that the formation rate of stars or GCs in galaxies is
proportional to the merger rate of galaxy halos (each galaxy undergoes
a major star burst episode when it virializes). Using the
Press-Schechter formalism I calculate,
\begin{eqnarray}
{d \omega^f_{gc}(z) \over dt}&=& {\cal A} \int_{M_1}^{M_2} {d \Omega(M_{DM}, z) \over
  dt} d\ln M_{DM},\label{eq:om1}\\
{d \omega^f_*(z) \over dt}&=& {\cal B} \int_{M_m}^\infty {d \Omega(M_{DM}, z) \over
  dt} d\ln M_{DM},
\label{eq:om}
\end{eqnarray}
where $\Omega(M_{DM},z)d\ln(M_{DM})$ is the mass fraction in
virialized DM halos of mass $M_{DM}$ at redshift z. I determine the
constants ${\cal A}$ and ${\cal B}$\footnote{By definition
$\int_0^\infty \Omega(M_{DM},z)d\ln(M_{DM})=1$. I find the following
values of the constants ${\cal A}=1.3\%, 1.6\%,0.6\%$ for cases (i),
(ii) and (iii) respectively (see text) and ${\cal B}=12\%$.} by
integrating eqs.~(\ref{eq:om1})-(\ref{eq:om}) with respect to time,
and assuming $\omega^f_{gc} = 0.1$\% (\ie, $f_{di}=2$) and
$\omega^f_*=1.4\omega_*=13$\% at $z=0$ (the factor 1.4 takes into
account the mass loss due to stellar winds and SN explosions adopting
a continuous star formation law).  I assume that GCs form in halos
with masses $M_1<M_{DM}<M_2$. The choice of $M_1$ and $M_2$ determine
the mean redshift, $z_f$, and time period, $\Delta t_{gc}$, for the
formation of old GCs. In order to be consistent with observations I
consider three cases: case (i) halos with virial temperature $2 \times
10^4 < T_{vir} < 5 \times 10^4$ K; case (ii) $5 \times 10^4 < T_{vir}
< 10^5$ K; and case (iii) $10^5 < T_{vir} < 5 \times 10^5$ K. In case
(i),(ii) and (iii) $\Delta t_{gc}=2.2, 3.7$ and 5.2 Gyr, respectively,
and the GC formation rate has a peak at $z = 7.5, 6$ and 4.6,
respectively. Disk and spheroid stars form in halos with
$M_{DM}>M_m$. At $z >10$ I assume that the first objects form in halos
with $M_m$ corresponding to a halo virial temperature $T_{vir}=5
\times 10^3$ K. At $z <10$ only objects with $T_{vir} > 2 \times 10^4$
K can form \citep[see][]{RicottiGSb:02}. The comoving star formation
rate, given by $\dot \rho_* = \overline \rho \dot \omega^f_*$, where
$\overline \rho = 5.51 \times 10^9$ M$_\odot$ Mpc$^{-3}$ is the mean
baryon density at $z=0$, is shown in Fig.~\ref{fig:sfr}. The points
show the observed SFR from \cite{Lanzetta:02}.
In Fig.~\ref{fig:nph} I show ${\cal N}_{ph}$ for GCs (thick lines)
and for galaxies (thin lines) defined as,
\begin{eqnarray}
{\cal N}^{gc}_{ph}&=& \eta f_{di}{d \omega^f_{gc} \over dt} t_{H}(z),\\
{\cal N}^*_{ph}&=& \eta \langle f_{esc}\rangle {d \omega^f_* \over dt}
t_{H}(z).
\end{eqnarray}
The thick solid, dashed and short-dashed lines show ${\cal N}^{gc}_{ph}$ for
case (i), (ii) and (iii), respectively. For comparison, I show (thin
solid line) ${\cal N}^*_{ph}$ assuming \fesc$=0.1 \times \exp[-z/2]$,
chosen to fit the observed values (squares) of ${\cal N}^*_{ph}$ at $z=2,3,4$
\citep{Miralda:00}. The thin dashed line shows ${\cal N}^*_{ph}$
assuming constant \fesc$=5$\%.

\section{Conclusions}\label{sec:disc}

Observed Lyman break galaxies at $z \sim 3$ are probably the most
luminous starburst galaxies of a population that produced the bulk of
the stars in our universe. Their formation epoch corresponds to the
assembly of the bulges of spirals and ellipticals. Nevertheless the
observed upper limit on \fesc from Lyman break galaxies is
\fesc$\simlt 10$\%, insufficient to reionize the IGM according to
numerical simulations.  Recently \cite{Ferguson:02}, using different
arguments, have claimed that the radiation emitted from Lyman
break galaxies is insufficient to reionize the IGM assuming a
continuous star formation mode.

I propose that GCs could produce enough ionizing photons to reionize
the IGM.  Assuming $f_{di}=2$ (\ie, during their evolution GCs have
lost half of their original mass), I find $\omega^f_{gc} \approx
0.1$\%, small compared to the total $\omega^f_* \sim 10$\% at $z=0$.
But GCs are around 12-13 Gyr old and, if they formed between $7<z<5$
(in about 0.5 Gyr), the expected total $\omega^f_*$ formed during this
time period is about $\omega^f_* \sim 1$\%, only 10 times larger than
$\omega^f_{gc}$. Assuming $f_{di}=20$, expected from the results of
N-body simulations, I find $\omega^f_{gc} \approx 1$\%, suggesting
that GC formation is an important mode of star formation at
high-redshift.  The special star formation mode required to explain
the formation of GCs suggests an \fesc$\sim 1$ from these objects.
This is because \fesc is dominated by the most luminous OB
associations and GCs are extremely luminous, emitting $S \sim 3 \times
10^{53}$ s$^{-1}$ ionizing photons in bursts lasting 4 Myrs.  Moreover,
according to many models, GCs form in the hot, collisionally-ionized
galaxy halo, from which all the ionizing radiation emitted can escape
into the IGM.  Therefore it is not too surprising, if GCs started
forming before $z = 6$, that their contribution to reionization is
large. I find that the number of ionizing photons per baryon emitted
in a Hubble time at $z=6$ by GCs is ${\cal
N}^{gc}_{ph}=(5.1^{+4.3}_{-3.2})f_{di}/\Delta t_{gc}>1$, therefore
sufficient to reionize the IGM even if we assume $f_{di}=1$. Here,
$\Delta t_{gc} \sim 0.5-2$ is the period of formation of the bulk of
old GCs in Gyrs. Using simple calculations based on Press-Schechter
formalism (see Fig.~\ref{fig:nph}) I find that, if normal star
formation in galaxies have \fesc$ \simlt 5$\%, GC contribution to
reionization should be important. If GCs formed by thermal instability
in the halo of $T_{vir} \sim 10^5$ K galaxies (case (iii)), the
ionizing sources are located in rare peaks of the initial density
field. Therefore, the mean size of intergalactic \HII regions before
overlap is large and reionization is inhomogeneous on large
scales.

In this letter I have considered the possibility that an increasing
\fesc at $z \sim 6$ due to GCs formation could explain IGM
reionization and still be consistent with the observed values of
the ionizing background at $z <3$. Alternatively an increasing
production of ionizing photons per baryon converted into stars, due to
a varying IMF, would have similar effects on the IGM. Chemical
evolution studies should be able to distinguish between these two
scenarios.


\bibliographystyle{/home/origins/ricotti/Latex/TeX/apj}
\bibliography{/home/origins/ricotti/Latex/TeX/archive}

\begin{thebibliography}{}

\bibitem[\protect\citeauthoryear{{Becker} et~al.}{{Becker}
  et~al.}{2001}]{Becker:01}
{Becker}, R.~H., et~al. 2001, \aj, 122, 2850

\bibitem[\protect\citeauthoryear{{Benson} et~al.}{{Benson}
  et~al.}{2002}]{Benson:02}
{Benson}, A.~J., {Lacey}, C.~G., {Baugh}, C.~M., {Cole}, S.,  \& {Frenk}, C.~S.
  2002, \mnras, 333, 156

\bibitem[\protect\citeauthoryear{{Bianchi}, {Cristiani}, \& {Kim}}{{Bianchi}
  et~al.}{2001}]{Bianchi:01}
{Bianchi}, S., {Cristiani}, S.,  \& {Kim}, T.-S. 2001, \aap, 376, 1

\bibitem[\protect\citeauthoryear{{Cayrel} et~al.}{{Cayrel}
  et~al.}{2001}]{Cayrel:01}
{Cayrel}, R., et~al. 2001, \nat, 409, 691

\bibitem[\protect\citeauthoryear{{Chaboyer} et~al.}{{Chaboyer}
  et~al.}{1998}]{Chaboyer:98}
{Chaboyer}, B., {Demarque}, P., {Kernan}, P.~J.,  \& {Krauss}, L.~M. 1998,
  \apj, 494, 96

\bibitem[\protect\citeauthoryear{{Chernoff} \& {Weinberg}}{{Chernoff} \&
  {Weinberg}}{1990}]{Chernoff:90}
{Chernoff}, D.~F.,  \& {Weinberg}, M.~D. 1990, \apj, 351, 121

\bibitem[\protect\citeauthoryear{{Ciardi}, {Bianchi}, \& {Ferrara}}{{Ciardi}
  et~al.}{2002}]{Ciardi:02}
{Ciardi}, B., {Bianchi}, S.,  \& {Ferrara}, A. 2002, \mnras, 331, 463

\bibitem[\protect\citeauthoryear{{Deharveng} et~al.}{{Deharveng}
  et~al.}{2001}]{Deharveng:01}
{Deharveng}, J.-M., {Buat}, V., {Le Brun}, V., {Milliard}, B., {Kunth}, D.,
  {Shull}, J.~M.,  \& {Gry}, C. 2001, \aap, 375, 805

\bibitem[\protect\citeauthoryear{{Djorgovski} et~al.}{{Djorgovski}
  et~al.}{2001}]{Djorgovski:01}
{Djorgovski}, S.~G., {Castro}, S., {Stern}, D.,  \& {Mahabal}, A.~A. 2001,
  \apjl, 560, L5

\bibitem[\protect\citeauthoryear{{Dove}, {Shull}, \& {Ferrara}}{{Dove}
  et~al.}{2000}]{Dove:00}
{Dove}, J.~B., {Shull}, J.~M.,  \& {Ferrara}, A. 2000, \apj, 531, 846

\bibitem[\protect\citeauthoryear{{Fall} \& {Rees}}{{Fall} \&
  {Rees}}{1985}]{Fall:85}
{Fall}, S.~M.,  \& {Rees}, M.~J. 1985, \apj, 298, 18

\bibitem[\protect\citeauthoryear{{Ferguson}, {Dickinson}, \&
  {Papovich}}{{Ferguson} et~al.}{2002}]{Ferguson:02}
{Ferguson}, H.~C., {Dickinson}, M.,  \& {Papovich}, C. 2002, \apjl, 569, L65

\bibitem[\protect\citeauthoryear{{Fukugita}, {Hogan}, \& {Peebles}}{{Fukugita}
  et~al.}{1998}]{Fukugita:98}
{Fukugita}, M., {Hogan}, C.~J.,  \& {Peebles}, P.~J.~E. 1998, \apj, 503, 518

\bibitem[\protect\citeauthoryear{{Giallongo} et~al.}{{Giallongo}
  et~al.}{2002}]{Giallongo:02}
{Giallongo}, E., {Cristiani}, S., {D'Odorico}, S.,  \& {Fontana}, A. 2002,
  \apjl, 568, L9

\bibitem[\protect\citeauthoryear{{Gnedin}}{{Gnedin}}{2000}]{Gnedin:00}
{Gnedin}, N.~Y. 2000, \apj, 535, 530

\bibitem[\protect\citeauthoryear{{Gnedin}}{{Gnedin}}{2002}]{Gnedin:02}
{Gnedin}, N.~Y. 2002, submitted (astro-ph/0110290)

\bibitem[\protect\citeauthoryear{{Gnedin}, {Lahav}, \& {Rees}}{{Gnedin}
  et~al.}{2002}]{GnedinR:02}
{Gnedin}, O.~Y., {Lahav}, O.,  \& {Rees}, M.~J. 2002, submitted
  (astro-ph/0108034)

\bibitem[\protect\citeauthoryear{{Gnedin} \& {Ostriker}}{{Gnedin} \&
  {Ostriker}}{1997}]{GnedinO:97}
{Gnedin}, O.~Y.,  \& {Ostriker}, J.~P. 1997, \apj, 474, 223

\bibitem[\protect\citeauthoryear{{Harris} \& {Pudritz}}{{Harris} \&
  {Pudritz}}{1994}]{Harris:94}
{Harris}, W.~E.,  \& {Pudritz}, R.~E. 1994, \apj, 429, 177

\bibitem[\protect\citeauthoryear{{Hu} et~al.}{{Hu} et~al.}{2002}]{Hu:02}
{Hu}, E.~M., {Cowie}, L.~L., {McMahon}, R.~G., {Capak}, P., {Iwamuro}, F.,
  {Kneib}, J.-P., {Maihara}, T.,  \& {Motohara}, K. 2002, \apjl, 568, L75

\bibitem[\protect\citeauthoryear{{Hurwitz}, {Jelinsky}, \& {Dixon}}{{Hurwitz}
  et~al.}{1997}]{Hurwitz:97}
{Hurwitz}, M., {Jelinsky}, P.,  \& {Dixon}, W. V.~D. 1997, \apjl, 481, L31

\bibitem[\protect\citeauthoryear{{Kundu} \& {Whitmore}}{{Kundu} \&
  {Whitmore}}{2001a}]{KunduI:01}
{Kundu}, A.,  \& {Whitmore}, B.~C. 2001a, \aj, 121, 2950

\bibitem[\protect\citeauthoryear{{Kundu} \& {Whitmore}}{{Kundu} \&
  {Whitmore}}{2001b}]{KunduII:01}
{Kundu}, A.,  \& {Whitmore}, B.~C. 2001b, \aj, 122, 1251

\bibitem[\protect\citeauthoryear{{Lanzetta} et~al.}{{Lanzetta}
  et~al.}{2002}]{Lanzetta:02}
{Lanzetta}, K.~M., {Yahata}, N., {Pascarelle}, S., {Chen}, H.,  \& {Fern{\'
  a}ndez-Soto}, A. 2002, \apj, 570, 492

\bibitem[\protect\citeauthoryear{{Larson}}{{Larson}}{1993}]{Larson:93}
{Larson}, R.~B. 1993, in ASP Conf. Ser. 48: The Globular Cluster-Galaxy
  Connection, 675

\bibitem[\protect\citeauthoryear{{Leitherer} et~al.}{{Leitherer}
  et~al.}{1999}]{Leitherer:99}
{Leitherer}, C., et~al. 1999, \apjs, 123, 3

\bibitem[\protect\citeauthoryear{{Madau} \& {Shull}}{{Madau} \&
  {Shull}}{1996}]{Madau:96}
{Madau}, P.,  \& {Shull}, J.~M. 1996, \apj, 457, 551

\bibitem[\protect\citeauthoryear{{McLaughlin}}{{McLaughlin}}{1999}]{McLaughlin%
:99}
{McLaughlin}, D.~E. 1999, \aj, 117, 2398

\bibitem[\protect\citeauthoryear{{McLaughlin} \& {Pudritz}}{{McLaughlin} \&
  {Pudritz}}{1996}]{McLaughlin:96}
{McLaughlin}, D.~E.,  \& {Pudritz}, R.~E. 1996, \apj, 457, 578

\bibitem[\protect\citeauthoryear{{Meiksin} \& {Madau}}{{Meiksin} \&
  {Madau}}{1993}]{Meiksin:93}
{Meiksin}, A.,  \& {Madau}, P. 1993, \apj, 412, 34

\bibitem[\protect\citeauthoryear{{Miralda-Escud\'e}, {Haehnelt}, \&
  {Rees}}{{Miralda-Escud\'e} et~al.}{2000}]{Miralda:00}
{Miralda-Escud\'e}, J., {Haehnelt}, M.,  \& {Rees}, M.~J. 2000, \apj, 530, 1

\bibitem[\protect\citeauthoryear{{Peebles} \& {Dicke}}{{Peebles} \&
  {Dicke}}{1968}]{Peebles:68}
{Peebles}, P. J.~E.,  \& {Dicke}, R.~H. 1968, \apj, 154, 891

\bibitem[\protect\citeauthoryear{{Ricotti}, {Gnedin}, \& {Shull}}{{Ricotti}
  et~al.}{2002a}]{RicottiGSa:02}
{Ricotti}, M., {Gnedin}, N.~Y.,  \& {Shull}, J.~M. 2002a, ApJ, submitted
  (Paper~I astro-ph/0110431)

\bibitem[\protect\citeauthoryear{{Ricotti}, {Gnedin}, \& {Shull}}{{Ricotti}
  et~al.}{2002b}]{RicottiGSb:02}
{Ricotti}, M., {Gnedin}, N.~Y.,  \& {Shull}, J.~M. 2002b, ApJ, submitted
  (Paper~II astro-ph/0110432)

\bibitem[\protect\citeauthoryear{{Ricotti} \& {Shull}}{{Ricotti} \&
  {Shull}}{2000}]{RicottiS:00}
{Ricotti}, M.,  \& {Shull}, J.~M. 2000, \apj, 542, 548

\bibitem[\protect\citeauthoryear{{Songaila} \& {Cowie}}{{Songaila} \&
  {Cowie}}{2002}]{Songaila:02}
{Songaila}, A.,  \& {Cowie}, L.~L. 2002, \aj, 123, 2183

\bibitem[\protect\citeauthoryear{{Steidel}, {Pettini}, \&
  {Adelberger}}{{Steidel} et~al.}{2001}]{Steidel:01}
{Steidel}, C.~C., {Pettini}, M.,  \& {Adelberger}, K.~L. 2001, \apj, 546, 665

\bibitem[\protect\citeauthoryear{{Stetson}, {Vandenberg}, \& {Bolte}}{{Stetson}
  et~al.}{1996}]{Stetson:96}
{Stetson}, P.~B., {Vandenberg}, D.~A.,  \& {Bolte}, M. 1996, \pasp, 108, 560

\bibitem[\protect\citeauthoryear{{Taniguchi}, {Trentham}, \&
  {Ikeuchi}}{{Taniguchi} et~al.}{1999}]{Taniguchi:99}
{Taniguchi}, Y., {Trentham}, N.,  \& {Ikeuchi}, S. 1999, \apjl, 526, L13

\bibitem[\protect\citeauthoryear{{van den Bergh}}{{van den
  Bergh}}{1999}]{vandenBergh:99}
{van den Bergh}, S. 1999, \aapr, 9, 273

\bibitem[\protect\citeauthoryear{{Wood} \& {Loeb}}{{Wood} \&
  {Loeb}}{2000}]{Wood:00}
{Wood}, K.,  \& {Loeb}, A. 2000, \apj, 545, 86

\bibitem[\protect\citeauthoryear{{Zepf} et~al.}{{Zepf} et~al.}{1999}]{Zepf:99}
{Zepf}, S.~E., {Ashman}, K.~M., {English}, J., {Freeman}, K.~C.,  \&
  {Sharples}, R.~M. 1999, \aj, 118, 752

\end{thebibliography}

\label{lastpage}
\end{document}